\documentclass[sigconf,review]{acmart}
\usepackage{amsmath,graphicx}
\usepackage{subfig}
\usepackage{multirow}
\usepackage{enumitem}
\usepackage{comment}
\usepackage{dblfloatfix}

\AtBeginDocument{%
  \providecommand\BibTeX{{%
    \normalfont B\kern-0.5em{\scshape i\kern-0.25em b}\kern-0.8em\TeX}}}

\copyrightyear{2021}
\acmYear{2021}
\setcopyright{usgovmixed}\acmConference[ICMI '21]{Proceedings of the 2021
International Conference on Multimodal Interaction}{October 18--22, 2021}{Montréal,
QC, Canada}
\acmBooktitle{Proceedings of the 2021 International Conference on Multimodal
Interaction (ICMI '21), October 18--22, 2021, Montréal, QC, Canada}
\acmPrice{15.00}
\acmDOI{10.1145/3462244.3481003}
\acmISBN{978-1-4503-8481-0/21/10}

\begin{document}

\title{Improved Speech Emotion Recognition using Transfer Learning and Spectrogram Augmentation}


\author{Sarala Padi}
\authornote{Both authors contributed equally to this research.}
\email{sarala.padi@nist.gov}
\affiliation{%
  \institution{ITL, NIST}
  \city{Gaithersburg}
  \state{MD}
  \country{USA}
}
\author{Seyed Omid Sadjadi}
\authornotemark[1]
\email{omid.sadjadi@nist.gov}
\affiliation{%
  \institution{ITL, NIST}
  \city{Gaithersburg}
  \state{MD}
  \country{USA}
}
\author{Ram D. Sriram}
\email{ram.sriram@nist.gov}
\affiliation{%
  \institution{ITL, NIST}
  \city{Gaithersburg}
  \state{MD}
  \country{USA}
}

\author{Dinesh Manocha}
\email{dmanocha@umd.edu}
\affiliation{%
  \institution{University of Maryland}
  \city{College Park}
  \state{MD}
  \country{USA}}

\renewcommand{\shortauthors}{Padi and Sadjadi, et al.}

\begin{abstract}
  Automatic speech emotion recognition (SER) is a challenging task that plays a crucial role in natural human-computer interaction. One of the main challenges in SER is data scarcity, i.e., insufficient amounts of carefully labeled data to build and fully explore complex deep learning models for emotion classification. This paper aims to address this challenge using a transfer learning strategy combined with spectrogram augmentation. Specifically, we propose a transfer learning approach that leverages a pre-trained residual network (ResNet) model including a statistics pooling layer from speaker recognition trained using large amounts of speaker-labeled data. The statistics pooling layer enables the model to efficiently process variable-length input, thereby eliminating the need for sequence truncation which is commonly used in SER systems. In addition, we adopt a spectrogram augmentation technique to generate additional training data samples by applying random time-frequency masks to log-mel spectrograms to mitigate overfitting and improve the generalization of emotion recognition models. We evaluate the effectiveness of our proposed approach on the interactive emotional dyadic motion capture (IEMOCAP) dataset. Experimental results indicate that the transfer learning and spectrogram augmentation approaches improve the SER performance, and when combined achieve state-of-the-art results. 
\end{abstract}




\maketitle

\section{Introduction}

Automatic emotion recognition plays a key role in human-computer interaction where it can enrich the next-generation AI with emotional intelligence by grasping the emotion from voice and words \cite{review-schuller,ser-intro-richard}. The motivation behind developing algorithms to analyze emotions is to design computer interfaces that mimic and embed realistic emotions in synthetically generated responses \cite{ser_intro_cowie}. Furthermore, research studies have shown that emotions play a critical role in the decision-making process for humans \cite{ser_intro_cowie}. Hence, there is a growing demand to develop automatic systems that understand and recognize human emotions. 

\begin{figure*}
\centering
 \includegraphics[scale=.5]{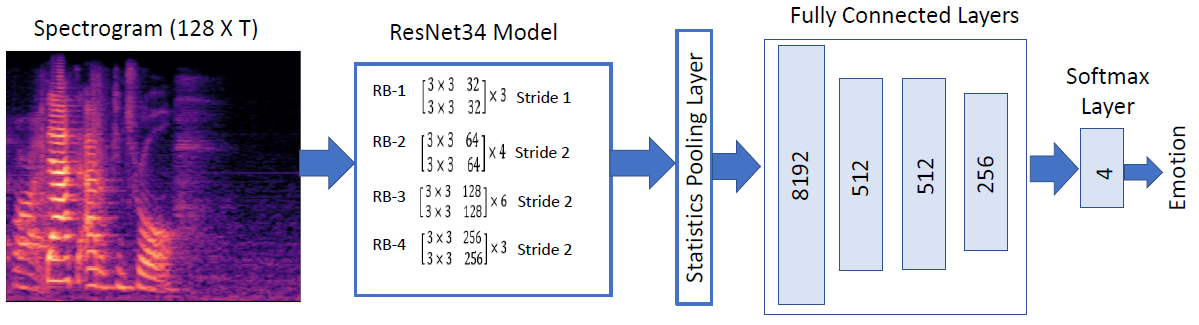}
   \caption{Block diagram of the proposed SER system. $T$ denotes the number of frames.}
\label{fig:method}
\end{figure*}

Humans express emotions in several ways, and speech is considered the most effective communication method to express feelings. For speech emotion recognition (SER), traditionally, machine learning (ML) models were developed using hand-crafted and engineered features such as mel-frequency cepstral coefficients (MFCC), Chroma-based features, pitch, energy, entropy, and zero-crossing rate \cite{intro-mfcc-ververidis,ser-dnn_han,intro-ml-kwon}, to mention a few. However, the performance of such ML models depends on the type and diversity of the features used. Although it remains unclear which features correlate most with various emotions, the research is still ongoing to explore additional features and new algorithms to model the dynamics of feature streams representing human emotions. On the other hand, the recent advancements in deep learning, along with the available computational capabilities, have enabled the research community to build end-to-end systems for SER. A big advantage of such systems is that they can directly learn the features from spectrograms or raw waveforms \cite{ CNN-adie-schuller, intro-wav-spect-yang,ser-spect-satt, ser-spect-ma, acm_cnngao}, thereby obviating the need for extracting a large set of hand-crafted features \cite{ser-tl-ghosh}. Recent studies have proposed the use of convolutional neural network (CNN) models combined with long short-term memory (LSTM) built on spectrograms and raw waveforms, showing improved SER performance \cite{ser-spect-satt,ser-spect-ma,ser-spect-yenigalla, Attentive_michael,CNN-RNN-schuller,spectral-mirsamadi,sarma2018emotion}. However, building such complex systems requires large amounts of labeled training data. Also, the insufficient labeled training data can potentially make the models overfit to specific data conditions and domains, resulting in poor generalization on unseen data.

This paper addresses the insufficient data problem using a transfer learning approach combined with a spectrogram augmentation strategy. We re-purpose a residual network (ResNet) model \cite{resnet} developed for speaker recognition using large amounts of speaker-labeled data and use it as a feature descriptor for SER. The model includes a statistics pooling layer that enables processing of variable length segments without a need for truncation. Also, we increase the training data size by generating more data samples using spectrogram augmentation \cite{spechaug}. We evaluate the effectiveness of our proposed system on the interactive emotional dyadic motion capture (IEMOCAP) dataset \cite{iemocap}. 

\section{Related work}

Recently, neural network based modeling approaches along with different variations of attention mechanism (e.g., plain \cite{huang2016emotion}, local \cite{spectral-mirsamadi}, and self \cite{table-tarantino2019self}) have shown promise for SER. Among them, techniques such as bidirectional LSTMs (BLSTM) \cite{huang2016emotion, spectral-mirsamadi, table-ramet2018context, table-tripathi, feng2020emotion} and time-delay neural networks (TDNN) \cite{wu2021bert}, which can effectively model relatively long contexts compared to their DNN counterparts, have been successfully applied for SER on the IEMOCAP. Nevertheless, as discussed previously, the lack of large amounts of carefully labeled data to build complex models for emotion classification remains a main challenge in SER \cite{acm_albanie2018emotion}. To address this, two approaches are commonly used: data augmentation and transfer learning. 

Data augmentation methods generate additional training data by perturbing, corrupting, mimicking, and masking the original data samples to enable the development of complex ML models. For example, \cite{sarma2018emotion, ser-gan-Aggelina, pappagari2020xvector} applied signal-based transformations such as speed perturbation, time-stretch, pitch shift, as well as added noise to original speech waveforms. One disadvantage of these approaches is that they require signal-level modifications, thereby increasing the computational complexity and storage requirements of the subsequent front-end processing. They can also lead to model overfitting due to potentially similar samples in the training set, while random balance can potentially remove useful information \cite{ser-gan-Aggelina}. For example, in \cite{ser-aug-caroline} a vocal tract length perturbation (VTLP) approach was explored for data augmentation along with a CNN model, and was found to result in a lower accuracy compared to a baseline model due to overfitting issues. 

Since generative adversarial network (GAN) based models have demonstrated remarkable success in computer vision, several studies have recently incorporated this idea to address the data scarcity problem and to generate additional data samples for SER~\cite{ser-gan-eskimez, ser-gan-Aggelina}. For instance, \cite{ser-gan-Aggelina} addressed the data imbalance using signal-based transformations and GAN based models for generating high-resolution spectrograms to train a VGG19 model for an emotion classification task and showed that GAN-generated spectrograms outperformed signal-based transformations. However, GAN generated features are strongly dependent on the data used during training and may not generalize to other datasets. Another challenge with GAN-based augmentation is that it is difficult to train and optimize. 

Another effective way to address challenges related to data scarcity is transfer learning~\cite{ser-tl-review-feng, gideon2017progressive, ottl2020group, boateng2020speech, zhao2021attention}. Transfer learning can leverage the information and knowledge learned from one related task and domain to another. Several recent studies have proposed transfer learning methods to improve SER performance and have shown these methods to outperform prior methods in recognizing emotions even for unseen scenarios, individuals, and conditions~ \cite{ser-tl-gideon}. It has been shown that transfer learning can increase feature learning abilities, and that the transferred knowledge can further enhance the SER classification accuracy \cite{ser-tl-ghosh,ser-tf-latif,ser-tf-schuller,ser-tf-song}. To further improve the SER performance, transfer learned features have been used in combination with deep belief networks (DBN) \cite{ser-tf-latif}, recurrent neural networks (RNN) \cite{ser-tl-ghosh}, CNN \cite{ser-tf-song}, temporal convolutional network (TCN) \cite{zhao2021attention}, and sparse autoencoder \cite{ser-tf-schuller}. However, transfer learning methods have not been fully explored and analyzed for emotion recognition. Particularly, it is unclear whether and how ML models trained for other data-rich speech applications such as speaker recognition would perform for SER.

\begin{figure*}[t]
\begin{center}
 \subfloat[\small {}] {\includegraphics[width=6cm,height=4cm]{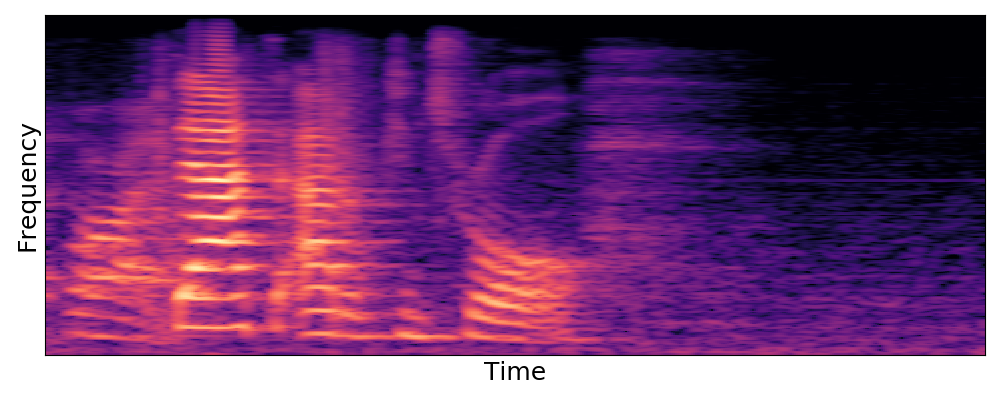}}
  \subfloat[\small {}] {\includegraphics[width=6cm,height=4cm]{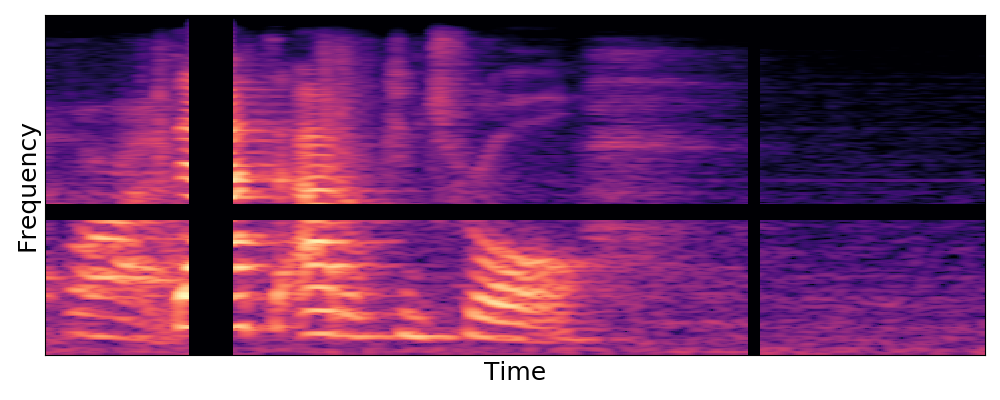}}
\end{center}
   \caption{\small{(a) original spectrogram, and (b) spectrogram modified using (multiple) masking blocks of consecutive time steps (vertical masks) and mel frequency channels (horizontal masks). The black horizontal and vertical stripes indicate the masked portions of the spectrogram.}}
\label{fig:spec}
\end{figure*}

\section{Proposed system}
\label{sec:system}
Figure~\ref{fig:method} shows the block diagram of the proposed system for speech emotion recognition. We use an end-to-end system with a ResNet34 model \cite{resnet} to perform emotion classification. ResNet models, originally developed for computer vision applications \cite{resnet}, have recently gained interest for speech applications such as speaker recognition \cite{zeinali2019but}. The residual blocks introduced in ResNet models allow us to train much deeper models that are otherwise difficult, if not impossible, to train due to vanishing and exploding gradient problems. The ResNet models also allow the higher layers to learn the identity function so that higher-level features perform equally well on unseen data compared to the lower layers of the model. In our proposed system, the convolutional layers in the model learn feature representations (feature maps) and reduce the spectral variations into compact representations, while the fully connected (FC) layers take the contextual features and generate predictions for emotion classification. 

\subsection{Input data}

Although SER systems traditionally used a large set of low-level time- and frequency-domain features to capture and represent the various emotions in speech, in recent years many state-of-the-art SER systems use complex neural network models that learn directly from spectrograms, or even raw waveforms. Accordingly, in this study, we build and explore a ResNet based system using log-mel spectrograms as input features. We extract high-resolution spectrograms to enable the model to not only learn the spectral envelope structure, but also the coarse harmonic structure for the various emotions.

\subsection{Transfer learning}

As noted previously, transfer learning is a ML method where a model initially developed for one task or domain is re-purposed, partly or entirely, for a different but related task/domain. It has recently gained interest for SER \cite{ser-tl-review-feng}. In this study, we re-purpose a model initially developed for speaker recognition to serve as a feature descriptor for SER. More specifically, we first train a ResNet34 model on large amounts of speaker-labeled audio data. Then, we replace the FC layers of the pre-trained model with new randomly initialized FC layers. Finally, we re-train the new FC layers for an SER task on the IEMOCAP dataset. 

\subsection{Statistics pooling}

As shown in Figure~\ref{fig:method}, the proposed system employs a statistics pooling layer \cite{snyder2018xvector} that aggregates the frame-level information over time and reduces the sequence of frames to a single vector by concatenating the mean and standard deviation computed over frames. Accordingly, the convolutional layers in the ResNet model work at the frame-level, while the FC layers work at the segment-level. This enables the system to efficiently model variable-length sequences of frames, thereby eliminating the need for truncating the sequence of frames to a pre-specified length to match that of the segments used during training. The sequence-truncation approach, which is commonly adopted in neural network based SER systems, can have a deleterious impact on SER performance as potentially informative frames are dropped out from the input. It is worth noting here that the statistics pooling can be viewed as an attention mechanism with equal weights for all frames, which also appends second order statistics (i.e., standard deviation) to capture long-term temporal variability over the duration of segments.  

\subsection{Spectrogram augmentation}

Currently, the majority of the features and methods for SER are adapted from speech recognition, speaker recognition, or speech synthesis fields \cite{ser-review}. There has been recent success in applying a computationally efficient data augmentation strategy, termed spectrogram augmentation, for speech recognition tasks \cite{spechaug}. The spectrogram augmentation technique generates additional training data samples by applying random time-frequency masks to spectrograms to mitigate the overfitting issue and improve the generalization of speech recognition models.  Motivated by promising results seen with the spectrogram augmentation in the speech recognition field, we augment the training data using spectro-temporally modified versions of the original spectrograms (see Figure~\ref{fig:spec}). Because the time-frequency masks are applied directly to spectrograms, the augmentation can be conveniently applied on-the-fly, eliminating the necessity to create and store new data files as commonly done in many augmentation approaches for speech applications.

\begin{table}[t]
    \caption{Parameter settings for the conservative and aggressive augmentation policies. Here, $N_f$ and $N_t$ denote the number of frequency and time masks applied.}
    \centering
    \begin{tabular}{|l|c|c|c|c|c|}
    \hline
         Augmentation Policy & $F$ & $W$ & $p$ & $N_f$ & $N_t$ \\ \hline\hline
         None & 0 & 0 & -- & -- & --   \\
         Conservative & 15 & 50 & 0.2 & 2 & 2  \\
         Aggressive  & 27 & 70 & 0.2 & 2 & 2  \\ \hline
    \end{tabular}
    
    \label{tab:specaug}
\end{table}

Similar to the approach taken in \cite{spechaug}, we consider two policies to systematically apply spectrogram augmentation for SER, namely conservative and aggressive. The frequency masking is applied over $f$ consecutive frequency channels in the range $\left[f_0,f_0+f\right)$, where $f$ is sampled from a uniform distribution $\left[0, F\right]$ and $f_0$ is sampled from $\left[0, \nu-f\right]$. Here, $F$ and $\nu$ denote the maximum width of frequency masks and the total number of frequency channels, respectively. The time masking, on the other hand, is applied over $t$ consecutive frames in the range $\left[t_0,t_0+t\right)$, where $t$ is selected from a uniform distribution $\left[0, W\right]$ and $t_0$ is sampled from $\left[0, T-t\right]$. Similarly, $W$ and $T$ denote the maximum width of time masks and the number of time frames, respectively. An upper bound is also applied on the width of the time masks such that $W=\min(W, pT)$, i.e., the width of a mask cannot be longer than $p$ times the number of time frames. This is to ensure sufficient speech content after masking, in particular for shorter segments. Table~\ref{tab:specaug} summarizes the various parameters for the two spectrogram augmentation policies used in this paper.  

\section{Experiments}
\label{sec:exp}
\subsection{Dataset}
We evaluate the effectiveness of the proposed SER system on the IEMOCAP dataset \cite{iemocap}, which contains improvised and scripted multimodal dyadic conversations between actors of opposite gender. It consists of 12 hours of speech data from 10 subjects, pre-segmented into short cuts that were judged by three annotators to generate emotion labels. It includes nine categorical emotions and 3-dimensional labels. In our experiments, we only consider the speech segments for which at least two annotators agree on the emotion label. In an attempt to replicate the experimental protocols used in a number of prior studies, we conduct three experiments on the full dataset (i.e., the combined improvised and scripted portions): Exp~1, using four categorical emotions: ``angry'', ``happy'', ``neutral'', ``sad''; Exp~2, using the same categories as in Exp~1, but replacing the ``happy'' category with ``excited''; Exp~3, by merging the ``happy'' and ``excited'' categories from Exp~1 and Exp~2. The total number of examples used for Exp~1 is $4490$ and the number of examples per category is $1103$, $595$, $1708$, and $1084$, respectively. The number of examples in the ``excited'' category is $1041$, making the total number of examples in the merged category (i.e., Exp~3) $1636$. Table~\ref{tab:statistics} summarizes the data statistics in the IEMOCAP dataset for the three experimental setups considered in this study. 

\begin{table}[t]
\caption{Data statistics for the various emotion classes in the IEMOCAP for the three experimental setups considered in this study. Both the improvised and scripted portions of the IEMOCAP dataset are used in our experiments.}
    \centering
    \begin{tabular}{|l|l|c|}
    \hline
         Experiment & Emotion & \#segments   \\
         \hline
         \hline
         \multirow{5}{*}{Exp 1} & Angry & 1103 \\
                                & Happy & 595 \\
                                & Neutral & 1708 \\
                                & Sad & 1084 \\\cline{2-3}
                                & \textbf{Total} & \textbf{4490} \\
                                \hline
         \multirow{5}{*}{Exp 2} & Angry & 1103 \\
                                & Excited & 1041 \\
                                & Neutral & 1708 \\
                                & Sad & 1084 \\\cline{2-3}
                                & \textbf{Total} & \textbf{4936} \\
                                \hline
         \multirow{5}{*}{Exp 3} & Angry & 1103 \\
                                & Excited+Happy & 1636 \\
                                & Neutral & 1708 \\
                                & Sad & 1084 \\\cline{2-3}
                                & \textbf{Total} & \textbf{5531} \\
                                \hline
    \end{tabular}
    
    \label{tab:statistics}
\end{table}

The IEMOCAP dataset comprises five sessions, and the speakers in the sessions are non-overlapping. Therefore, there are 10 speakers in the dataset, i.e., 5 female and 5 male speakers. To conduct the experiments in a speaker-independent fashion, we use a leave-one-session-out (LOSO) cross-validation strategy, which results in 5 different train-test splits/folds. For each fold, we use the data from 4 sessions for training and the remaining one session for model evaluation. Since the dataset is multi-label and imbalanced, in addition to the overall accuracy, termed weighted accuracy (WA), we report the average recall over the different emotion categories, termed unweighted accuracy (UA), to present our findings. Additionally, to understand and visualize the performance of the proposed system within and across the various emotion categories, we compute and report confusion matrices for the three experiments. Note that for each experiment, we compute the average of performance metrics over the five training-test splits as the final result.

\subsection{Setup and configuration}\label{subsec:exp}

For speech parameterization, high resolution 128-dimensional log-mel spectrograms are extracted from $25$~ms frames at a $100$~Hz frame rate (i.e., every $10$~ms). For feature normalization, a segment level mean and variance normalization is applied\footnote{No voice activity detection (VAD) is applied prior to feature normalization because it was found to be detrimental to SER performance on the IEMOCAP. We hypothesize that this is because the silence gaps in within and between utterances might be relevant in terms of speakers' emotional state.}. Note that this is not ideal as typically the normalization is applied at the recording/conversation level. We have found that normalizing the segments using statistics computed at the conversation level significantly improves the SER performance on the IEMOCAP. Nevertheless, this violates the independence assumption for the speech segments, hence it is not considered in this study. The front-end processing, including feature extraction and feature normalization, is performed using the NIST speaker and language recognition evaluation (SLRE) \cite{sadjadi2020sre, sadjadi2018lre} toolkit. While training the model, we select $T$-frame chunks using random offsets over original speech segments where $T$ is randomly sampled from the set $\{150, 200, 250, 300\}$ for each batch. For speech segments shorter than $T$ frames, signal padding is applied. On the other hand, while evaluating the model, we feed the entire duration of the test segments because the statistics pooling layer enables the model to consume variable-length inputs. 

\begin{table*}[t]
\caption{Performance comparison of our proposed approach with prior methods that use the LOSO strategy for experiments on the full IEMOCAP dataset (i.e., both the improvised and scripted portions).  Abbreviations: A-Angry, H- Happy, N-Neutral, E-Excited, H+E: Happy and Excited merged. Blanks (--) indicate unreported values.}
\label{tab:Comp_2}
\centering
\begin{tabular}{|l|l|c|c|} 
 \hline
{Experiment (emotion classes)} & {Approach} & {UA [\%]} & {WA [\%]} \\  \hline\hline
\multirow{5}{*}{Exp 1 (A, H, S, N)}  
                                   & BLSTM+attention \cite{huang2016emotion} & 49.96 & 59.33 \\
                                   & Transformer+self-attention \cite{table-tarantino2019self} & 58.01 & 59.43 \\
                                   & BLSTM+local attention \cite{spectral-mirsamadi} & 58.8  & 63.5 \\
                                   & BLSTM+attention \cite{table-ramet2018context} & 59.6 & 62.5 \\
                                   \cline{2-4}
                                   & {\bf Proposed} & {\bf 61.61} & {\bf 66.02} \\\hline
\multirow{4}{*}{Exp 2 (A, E, S, N)}  & CTC-BLSTM \cite{table-chernykh2017emotion} & 54 & -- \\
                                   & BLSTM+attention \cite{table-tripathi} & 55.65 & --  \\
                                   & Transformer+self-attention \cite{table-tarantino2019self}           & 64.79 & 64.33 \\\cline{2-4}
                                   & {\bf Proposed} & {\bf 65.56} & {\bf 65.62} \\\hline
            
\multirow{7}{*}{Exp 3 (A, H+E, S, N)}
                                   & BLSTM+transfer learnig \cite{ser-tl-ghosh} & 51.86 & 50.47 \\
                                   & VGG19+GAN augmentation \cite{ser-gan-Aggelina} & 54.6 & -- \\
                                   & CNN+attention+multi-task learning \cite{Attentive_michael} & -- & 56.10 \\
                                   & BLSTM+self-attention \cite{feng2020emotion} & 57.0 & 55.7 \\
                                   & CNN+attention+multi-task/transfer learning \cite{neumann2019attentive}           & 59.54 & -- \\
                                   & ResTDNN+self-attention \cite{wu2021bert} & 61.32 & 60.64 \\
                                   \cline{2-4}
                                   & {\bf Proposed} & {\bf 64.14} & {\bf 63.61} \\\hline
            
\end{tabular}
\end{table*}

As noted previously, the proposed end-to-end SER system uses a pre-trained ResNet34 model built on a speaker recognition task. We train the ResNet34 model on millions of speech samples from more than 7000 speakers available in the VoxCeleb corpus~\cite{nagrani2020voxceleb}. To build the speaker recognition model, we apply the same front-end processing described above to extract high-resolution log-mel spectrograms from VoxCeleb data. We conduct experiments using models with and without transfer learning and spectrogram augmentation. For each original speech segment, we generate and augment two spectro-temporally modified versions according to the augmentation policies defined in Table~\ref{tab:specaug}. This is applied for both speaker and emotion recognition systems during training. To study the impact of the statistics pooling layer, we also evaluate these models with and without this layer. For all the experiments, we use a categorical cross-entropy loss as the objective function to train the models. The number of channels in the first block of the ResNet model is set to $32$. The model is trained using Pytorch\footnote{https://github.com/pytorch/pytorch} and the stochastic gradient descent (SGD) optimizer with momentum ($0.9$), an initial learning rate of $10^{-2}$, and a batch size of $32$. The learning rate remains constant for the first $8$ epochs, after which it is halved every other epoch. We use parametric rectified linear unit (PReLU) activation functions in all layers (except for the output), and utilize a layer-wise batch normalization to accelerate the training process and improve the generalization properties of the model.      

\begin{figure*}[t]
\begin{center}
\subfloat[\small {Exp~1: A, H, N, S }]{\includegraphics[scale=0.45]{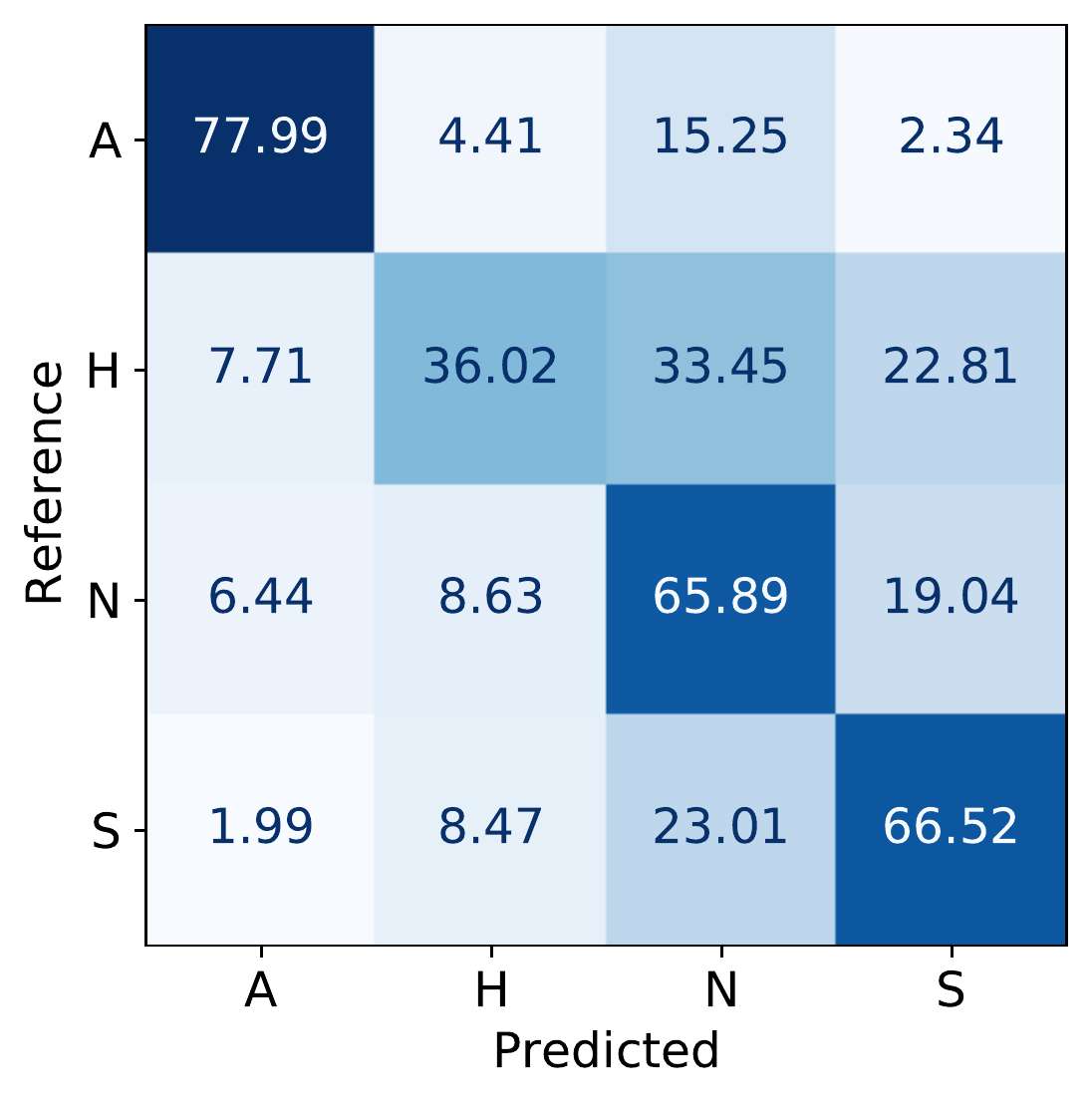}}
\hspace{0.5cm}
 \subfloat[\small {Exp~2: A, E, N, S }]{\includegraphics[trim=.7cm 0cm 0cm 0cm, clip=true, scale=0.45]{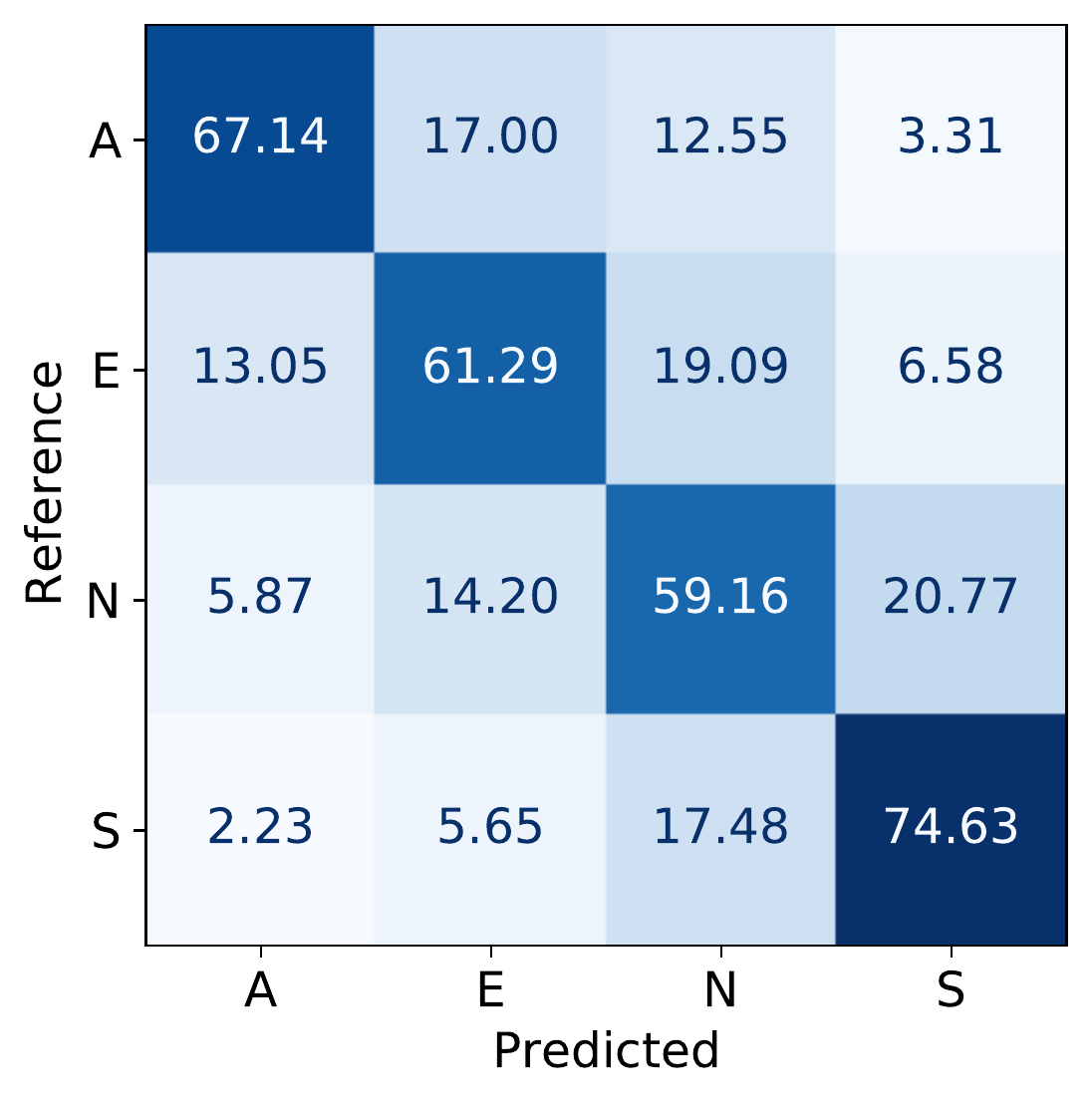}}
 \hspace{0.3cm}
 \subfloat[\small {Exp~3: A, H+E, N, S }]{\includegraphics[trim=.7cm 0cm 0cm 0cm, clip=true,scale=0.45]{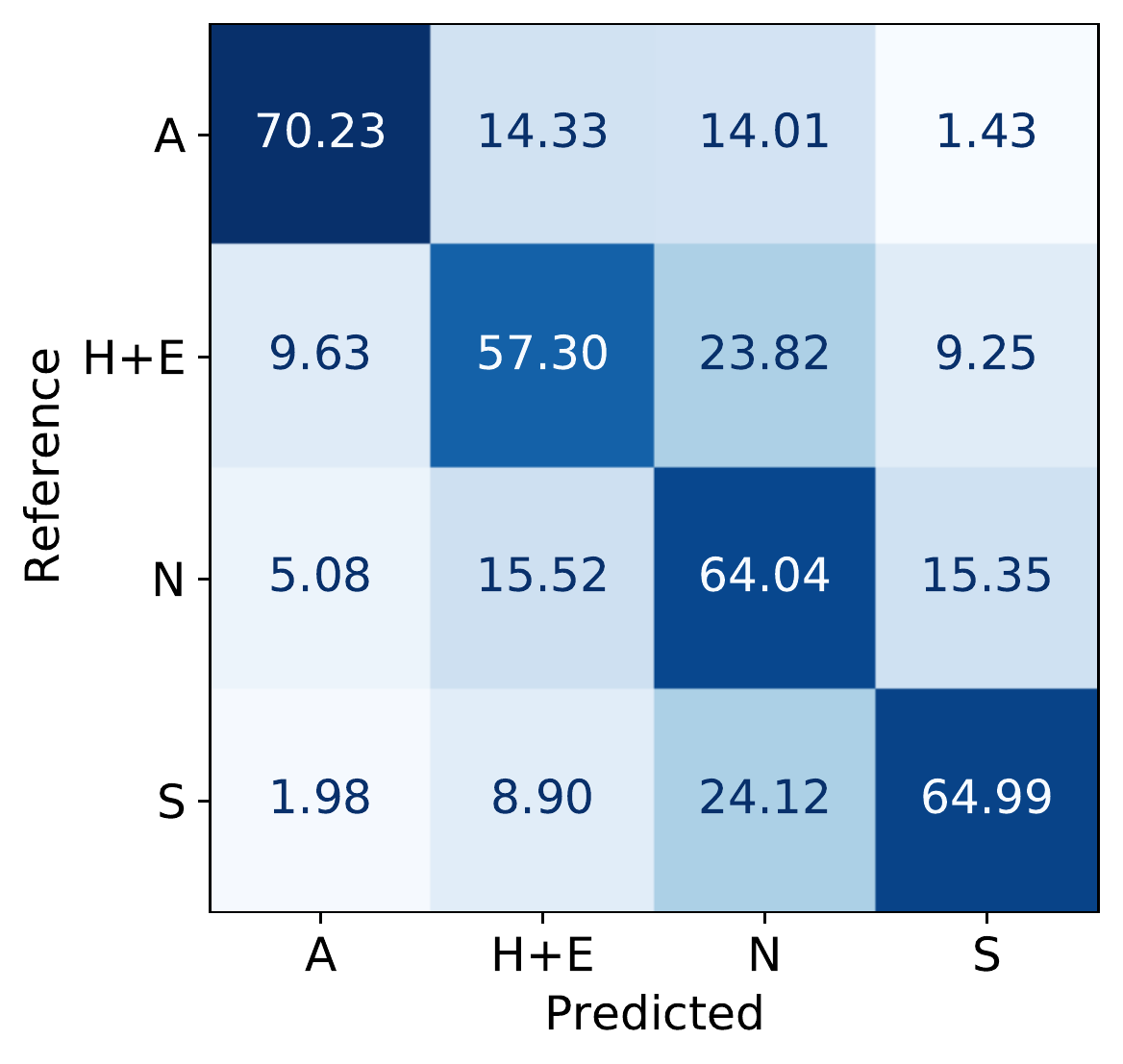}}\\
\end{center}
   \caption{ Performance confusion matrices of the proposed SER system for the three experiments conducted in this study using the LOSO strategy. Abbreviations: A-angry, E-excited, H-happy, N-neutral, S-Sad, and H+E- Happy and Excited merged. }
\label{fig:confusion}
\end{figure*}

\begin{figure*}[t]
 \includegraphics[scale=.5]{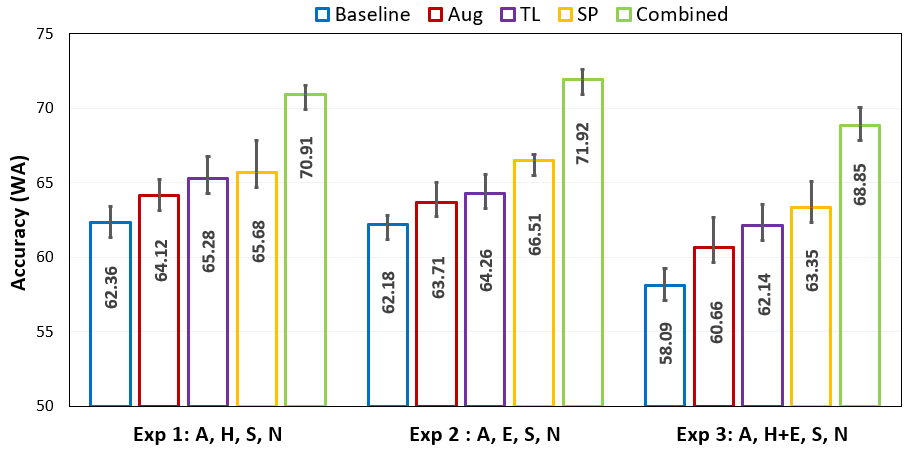}
   \caption{Performance (WA) of the proposed approach with and without transfer learning (TL), spectrogram augmentation (Aug), and statistics pooling (SP). All results are obtained using a 5-fold cross-validation. The height of each bar represents the average accuracy computed over 5 runs, while the error bars denote the standard deviations over the 5 runs. Abbreviations: A-Angry, H-Happy, N-Neutral, S-Sad, E-Excited, H+E: Happy and Excited merged}
\label{fig:method}
\vspace{0.5cm}
\end{figure*}

\section{Results}

Table~\ref{tab:Comp_2} presents the performance comparison of our proposed system with several prior approaches for the three experimental setups (i.e., Exp~1, 2, and 3) described in Section~\ref{sec:exp}. The results are obtained using the combined system that utilizes the ResNet model with the statistics pooling layer trained using the transfer learning and spectrogram augmentation approaches described in Section~\ref{sec:system}. All studies referenced in the table adopt the LOSO strategy to conduct experiments on both the improvised and scripted portions of the IEMOCAP dataset\footnote{There are other related studies in the literature that only use the improvised portion of the IEMOCAP dataset \cite{ser-aug-caroline, ser-spect-satt,zhao2021attention}. On the other hand, in our experiments, we use both the improvised and scripted portions of the IEMOCAP, which is approximately twice the size of the improvised portion alone. Because the experimental setups and the amount of data used for model training and evaluation in those studies are different than ours, we have not included them in Table~\ref{tab:Comp_2} for comparison. The SER performance on the improvised portion is known to be better than that on the full dataset (e.g., see \cite{table-tripathi, table-ramet2018context, table-tarantino2019self, ser-tl-ghosh, Attentive_michael}).}.
It can be seen from the table that the proposed system consistently provides competitive performance across the three experiments, achieving state-of-the-art results. In the case of Exp~2, the proposed system outperforms a system that uses $384$ engineered features \cite{table-tarantino2019self}, while for the other two experiments, our proposed system outperforms systems that use a large set of engineered features (e.g., \cite{spectral-mirsamadi} and \cite{table-ramet2018context}).


To visualize the performance of the proposed system within and across the different emotion categories, confusion matrices for the three experimental setups are shown in Figure~\ref{fig:confusion}. It is observed from Figure~\ref{fig:confusion}(a) that the system confuses the ``happy'' class (H) with the ``neutral'' class (N) quite often, while performing the best on the ``angry'' class (A). This is consistent with observations reported in other studies on IEMOCAP \cite{Attentive_michael, yoon2018multimodal}. Our informal listening experiments confirm that the ``happy'' and ``neutral'' classes are indeed confusable emotion pairs in the IEMOCAP dataset. The system performance balance is improved in Figure~\ref{fig:confusion}(b) where we replace the less pronounced ``happy'' category with the ``excited'' category (E). Combining the ``happy'' and ``excited'' categories in Exp~3 further improves the performance balance across the various emotions, at the expense of increasing the confusion between the ``angry'' (A) and ``excited'' plus ``happy'' (H+E) categories. 

To investigate and quantify the contribution of the various system components proposed in this study for improved SER, we further conduct ablation experiments to measure the system performance with and without the transfer learning, the spectrogram augmentation, and the statistics pooling layer. For these ablation experiments, we employ a 5-fold cross-validation (CV) strategy, where we use 80\% of the data for training and 20\% for testing the system. This process is repeated 5 times to reduce possible partition-dependencies. Figure~\ref{fig:method} shows the average overall classification accuracy (WA) computed across 5 folds (or 5 runs). The height of the bars represents the average accuracy, and the error bars denote the standard deviations computed over the 5 runs. It is observed that the proposed components, both individually and in combination, consistently provide performance gains across the three experimental setups (i.e., Exp~1, 2 and 3). The statistics pooling approach seems to have the greatest impact on performance, followed by the transfer learning and spectrogram augmentation methods. Furthermore, the model that combines all the system components not only consistently achieves the best performance, but also relatively smaller variation across the 5 runs as evidenced by the error bars.  

\section{Conclusions}
In this paper, we explored a transfer learning approach along with a spectrogram augmentation strategy to improve the SER performance. Specifically, we re-purposed a pre-trained ResNet model from speaker recognition that was trained using large amounts of speaker-labeled data. The convolutional layers of the ResNet model were used to extract features from high-resolution log-mel spectrograms. In addition, we adopted a spectrogram augmentation technique to generate additional training data samples by applying random time-frequency masks to log-mel spectrograms to mitigate overfitting and improve the generalization of emotion recognition models. We evaluated the proposed system using three different experimental settings and compared the performance against that of several prior studies. The proposed system consistently provided competitive performance across the three experimental setups, achieving state-of-the-art results on two settings. The state-of-the-art results were achieved without the use of engineered features. It was also shown that incorporating the statistics pooling layer to accommodate variable-length audio segments improved the emotion recognition performance. Results from this study suggest that, for practical applications, simplified front-ends with only spectrograms can be as effective for SER, and that models trained for data-rich speech applications such as speaker recognition can be re-purposed using transfer learning to improve the SER performance under data scarcity constraints.  
In the future, to further enhance the emotion recognition accuracy, we will extend our work along these lines by exploring more data augmentation methods, incorporating other transfer learning paradigms, and evaluating the proposed system across different datasets.

\section{Acknowledgement}
Experiments and analyses were performed, in part, on the NIST Enki HPC cluster.

\section{Disclaimer}
The views and conclusions presented in this paper are those of the authors and should not be interpreted as the official findings, either expressed or implied, of NIST or the U.S. Government.

\bibliographystyle{ACM-Reference-Format}
\bibliography{sample-base}


\end{document}